\documentclass[letterpaper, 10 pt, conference]{ieeeconf}  

\IEEEoverridecommandlockouts                              

\usepackage{graphicx} 
\usepackage{amsmath} 
\usepackage{amssymb}  
\usepackage{float}
\usepackage{booktabs}
\usepackage{tensor}
\usepackage{accents}
\usepackage[ruled,vlined]{algorithm2e}
\usepackage{subcaption}
\usepackage{mathtools}
\usepackage{textcomp}
{
    \newtheorem{assumption}{Assumption}
}

\newcommand{\ignore}[1]{}

\newcommand{\bma}[1]{\left[\begin{array}{#1}}
\newcommand{\ema}{\end{array}\right]}

\DeclareMathAlphabet{\mbf}{OT1}{ptm}{b}{n}

\def\fdotb{{\raisebox{-0.6ex}{ \kern0.2ex\raisebox{0.8ex}{\tiny $\hspace*{-1ex}\circ$}}}}
\def\fddotb{{\raisebox{-0.6ex}{ \kern0.2ex\raisebox{0.8ex}{\tiny $\hspace*{-1ex}\circ\circ$}}}}

\newcommand{\utimes}{ {\raisebox{-0.6ex}{ \kern-1.0ex\raisebox{0.6ex}{ \small $\mathsf{v}$}}} } %
\newcommand{\beq}{\begin{equation}}
\newcommand{\eeq}{\end{equation}}
\newcommand{\bdis}{\begin{displaymath}}
\newcommand{\edis}{\end{displaymath}}
\newcommand{\beqarray}{\begin{eqnarray}}
\newcommand{\eeqarray}{\end{eqnarray}}
\newcommand{\beqarraynn}{\begin{eqnarray*}}
\newcommand{\eeqarraynn}{\end{eqnarray*}}

\SetKw{Break}{break}

\title{\LARGE \bf Robust Output Feedback of Nonlinear Systems through the Efficient Solution of Min-Max Optimization Problems}

\author{Jad Wehbeh$^{1}$ and Eric C. Kerrigan$^{2}$
\thanks{$^{1}$Jad Wehbeh is with the Department of Electrical and Electronic Engineering, Imperial College London, SW7 2AZ, UK
        {\tt\small j.wehbeh22@imperial.ac.uk}}%
\thanks{$^{2}$Eric C. Kerrigan is with the Department of Electrical and Electronic Engineering and the Department of Aeronautics, Imperial College London,
        SW7 2AZ, UK
        {\tt\small e.kerrigan@imperial.ac.uk}}
}

\begin{document}

\maketitle
\thispagestyle{empty}
\pagestyle{empty}

\begin{abstract}
    We examine robust output feedback control of discrete-time nonlinear systems with bounded uncertainties affecting the dynamics and measurements. Specifically, we demonstrate how to construct semi-infinite programs that produce gains to minimize some desired performance cost over a finite prediction horizon for the worst-case realization of the system's uncertainties, while also ensuring that any specified nonlinear constraints are always satisfied. The solution process relies on an implicit description of the feasible state space through prior measurements and the system dynamics, and assumes that the system is always in the subset of the feasible space that is most detrimental to performance. In doing so, we can guarantee that the system's true state will meet all of the chosen performance criteria without resorting to any explicit state estimation. Under some smoothness assumptions, we also discuss solving these semi-infinite programs through local reduction techniques, which generate optimal scenario sets for the uncertainty realizations to approximate the continuous uncertainty space and speed up the computation of optima. When tested on a two-dimensional nonlinear quadrotor, the developed method achieves robust constraint satisfaction and tracking despite dealing with highly uncertain measurements and system dynamics.
\end{abstract}


\section{Introduction}

\subsection{Background and Motivation}

The introduction of the separation principle in the early 1960s fundamentally altered the approach to controller design, decoupling the control and estimation subproblems for a large class of linear systems \cite{joseph1961linear}. Ever since, it has been widely accepted that the guidance and control of any uncertain system should proceed through the two standard phases of state estimation, where noisy measurements are converted into a higher-fidelity version of the system's true state, and state feedback control, where this state estimate is used to generate a control input sequence that directs the system along some desired trajectory \cite{aastrom2021feedback}.

Unfortunately, the separation principle does not apply to uncertain nonlinear systems in general, which are usually \textit{dual}~\cite{feldbaum1963dual}. This means that the control trajectories chosen for these systems directly affect the quality of the information produced by the measurements, effectively coupling any estimation subproblem with the control design. As a consequence, it has been shown that the stability of a control law applied to a general nonlinear system cannot be guaranteed, since the controller does not take into account the possible interactions with the measurements obtained~\cite{feldbaum1963dual}. It is also well understood that for nonlinear systems, condensing output sequences into box bounds on the states, or Gaussian estimates of the means and associated covariances, sacrifices information when compared to the original measurements~\cite{barfoot2024state}. This is a direct result of Gaussian distributions and box constraints not being preserved by nonlinear operations, rendering any such estimate an approximation of the true resulting distribution. 

Output feedback has emerged as a possible solution to these issues, forsaking the traditional state estimation component of the control problem. Instead, the control law maps the measurements directly onto the desired control inputs. Provided that the system is modeled appropriately, this facilitates the description of the interactions between the control inputs produced and the expected measurements. This also allows for an implicit description of the state evolution through a history of past measurements, avoiding any of the shortfalls associated with explicit state estimation.

\subsection{State of the Art}

Static output feedback is perhaps the simplest form of direct control input generation from output measurements. This family of approaches generates the inputs $u$ from the system outputs or measurements $y$ using the linear feedback law $u = -Ky$, where $K$ is a control gain that can be designed to ensure desired system properties~\cite{syrmos1997static}. Most applications of static output feedback focus on linear time-invariant (LTI) systems, and usually attempt to guarantee Lyapunov stability by obtaining $K$ through the solution of tailored linear matrix inequalities (LMIs). Some more modern static output feedback techniques attempt to achieve better robustness properties by decoupling the gains from the Lyapunov matrices, but generally remain limited in the systems they can stabilize \cite{sadabadi2016static}.

Dynamic output feedback extends this idea of direct feedback on the output to a controller with memory of past measurements and time-varying gains \cite{amato2006finite}. The controller is usually modeled as its own discrete-time linear system, and the gains at each time step are once again obtained by solving an appropriate set of LMIs \cite{deaecto2011dynamic}. Most applications of dynamic output feedback consider LTI systems, but more recent work also examined switched systems with delays \cite{yang2019output} or problems with nonlinear dynamics represented through Takagi-Sugeno (T-S) fuzzy systems \cite{liu2016dynamic, zhang2023improved}. Model predictive control (MPC) formulations using output feedback have also been suggested for linear \cite{goulart2007output} and linear parameter varying (LPV)~\cite{de2021robust} systems with bounded disturbances and measurement noise, as well as stochastic systems with zero-mean white noise \cite{farina2015approach}. However, none of the existing MPC or dynamic output feedback techniques are capable of guaranteeing robustness for arbitrary nonlinear systems subject to multiplicative or more general uncertainties.

Other relevant examples of direct output feedback control can be found in the literature on data-driven control and reinforcement learning. Although a large portion of the work on classical model-free control assumes access to direct state measurements or noise-free output measurements, some recent approaches consider data-driven output feedback control on noisy data \cite{tang2022data}. For example, data-enabled predictive control (DeePC) is capable of handling linear systems with white measurement noise \cite{coulson2019data}, and its robust extension deals with a wide class of bounded uncertainties in the input-output mapping by solving a min-max problem to compute the desired input sequence \cite{huang2023robust}. The work of~\cite{gong2023data} also provides some interesting perspectives on data-driven output feedback for systems with disturbances and uncertainties, with a focus on applications in power grids. Nevertheless, these data-driven methods remain limited in the scope of systems to which they are applicable, and trade dependence on knowledge of the model for the requirement of sufficiently rich data sequences to initialize the controller.

Reinforcement learning approaches, meanwhile, are capable of learning the input-output mapping for a wide variety of nonlinear and uncertain systems, allowing direct input feedback for a far larger range of systems than any of the other methods discussed \cite{nian2020review,kiran2021deep}. However, these methods generally require access to large amounts of data and repeated testing to obtain high quality approximations of the input-output mapping, and commonly fail to provide any guarantees on the robustness of the strategies produced. Safe reinforcement learning attempts to address the latter of these shortcomings, but is somewhat limited in the systems to which it can be applied~\cite{brunke2022safe}. 

\subsection{Contributions}

In this paper, we demonstrate how the problem of computing output feedback gains that optimize worst-case performance for uncertain nonlinear systems can be described as a semi-infinite program of the kind discussed in \cite{wehbeh2024efficient}, and solved using the local reduction methods for dynamic optimization described in \cite{blankenship1976infinitely} and \cite{zagorowska2024automatic}. While doing so, we introduce the novel contributions of:
\begin{itemize}
    \item Implicitly describing the feasible state space of the system as a function of the measurements using constraints derived from the dynamics and measurement equations.
    \item Demonstrating how this description of the state-space can be used to define semi-infinite programs that design optimal robust gains for output feedback. The gains obtained through this approach ensure that a desired level of performance in the state can be achieved without explicitly relying on any state estimation.  
    \item Showcasing the performance of the proposed approach on a two-dimensional quadrotor tracking problem.
\end{itemize}


\section{System Description}
\label{sec:system}

Consider a general nonlinear system with discrete-time dynamics described by
\begin{equation}
    \label{eqn:dyn}
    f(x_k,x_{k+1},u_k,\rho_f,w_k) = 0
\end{equation}
where $x_k \in \mathcal{X} \subseteq \mathbb{R}^{n_x}$ is the state of the system at time step~$k$, $u_k \in \mathcal{U} \subseteq \mathbb{R}^{n_u}$ represents the control input applied at step~$k$, $\rho_f \in \mathcal{R}_f \subseteq \mathbb{R}^{n_f}$ describes any unknown parameters in $f$, and $w_k \in \mathcal{W} \subseteq \mathbb{R}^{n_w}$ is a vector of time-varying uncertainties or disturbances affecting the state evolution at step~$k$. 

\begin{assumption}
    \label{assum:f_continuity}
    The function $f(\cdot,\cdot,\cdot,\cdot,\cdot):$ $\mathcal{X} \times \mathcal{X} \times \mathcal{U} \times \mathcal{R}_f \times \mathcal{W} \rightarrow \mathbb{R}^n$ is continuous in all its arguments, i.e.\  $f(\cdot,\cdot,\cdot,\cdot,\cdot) \in \mathcal{C}^{\,0}$.
\end{assumption}

As with most physical systems, it is impossible to know the true value of $x_k$ at any given time. Instead, at each time step $k$, we obtain the measurement $y_k$ through the measurement equation
\begin{equation}
    \label{eqn:measurements}
    y_k = h(x_k, u_k, \rho_h, v_k)
\end{equation}
where $\rho_h \in \mathcal{R}_h \subseteq \mathbb{R}^{n_h}$ describes any unknown parameters in $h$, and $v_k \in \mathcal{V} \subseteq \mathbb{R}^{n_v}$ is a vector of time-varying uncertainties or measurement noises affecting the measured output at step~$k$.
\begin{assumption}
    The function $h(\cdot,\cdot,\cdot,\cdot):$ $\mathcal{X} \times \mathcal{U} \times \mathcal{R}_h \times \mathcal{V} \rightarrow \mathcal{Y}$ is continuous in all of its arguments, i.e.\  $h(\cdot,\cdot,\cdot,\cdot) \in \mathcal{C}^{\,0}$.
\end{assumption}
\begin{assumption}
    \label{assum:noise_compactness}
    Even though $\rho_f$, $\rho_h$, $w_k$, and $v_k$ cannot be known directly, the sets $\mathcal{R}_f$, $\mathcal{R}_h$, $\mathcal{W}$, and $\mathcal{V}$ are compact and known. This implies that all uncertainties affecting the system are bounded with known bounds.
\end{assumption}

Note that while obtaining the true bounds on the noise  for Assumption \ref{assum:noise_compactness} may seem difficult to achieve in practice, the choice of sets for $\mathcal{R}_f$, $\mathcal{R}_h$, $\mathcal{W}$, and $\mathcal{V}$ need not be tight, and must simply contain all possible values of the uncertainties. Using larger uncertainty bounds than the true values only leads to more conservative performance. Disturbances affecting physical systems are generally limited by the total amount of energy available, making it possible to come up with bounds on the uncertainties and compact set descriptions for the possible realizations.

Next, we consider the feedback law used to generate the control inputs $u_k$. Given that this work focuses on output feedback, we do not assume access to the true values of the state or the output of a state estimator, deriving our feedback instead from measurements of $y_k$, as well as measurements from the previous $M$ time steps, to obtain the control law 
\begin{equation}
    u_k = \phi_k(\theta_k,Y^-_M)
\end{equation}
where $\theta_k \in \varTheta \subseteq \mathbb{R}^{n_\theta}$ represents the gains and parameters used in generating $u_k$ and
\begin{equation*}
    Y^-_M := (y_{k-M}, y_{k-M+1}, \hdots, y_{k-1}, y_k)
\end{equation*}
is the collection of measurements between $k-M$ and $k$. 

The goal of this work is therefore to design a sequence~of $\theta_k$ over a fixed prediction horizon of $N$ time steps that minimizes some objective
\begin{equation*}
    J_k \left( X^+_N,\bar{X}^+_N,Y^+_N, \bar{Y}^+_N,U^+_N \right)
\end{equation*}
where $\bar{x}_k \in \mathcal{X}$ is a reference value for $x_k$ and $\bar{y}_k \in \mathcal{Y}$ is a reference value for $y_k$, and where
\begin{equation*}
    X^+_N := (x_{k+1}, x_{k+2}, \hdots, x_{k+N-1}, x_{k+N}),
\end{equation*}
with $\bar{X}^+_N$, ${Y}^+_N$, and $\bar{Y}^+_N$ similarly defined. $U^+_N$, meanwhile, is defined between the time steps $k$ and $k+N-1$. 

Although this is not something we examine explicitly in this paper, we allow $J_k$ to depend on $y$ and $\bar{y}$ to facilitate the modelling of problems that target estimation objectives or express their goal in terms of measurement tracking instead of state tracking. 

Given that the value taken by $J_k$ will also depend on the uncertainties $\rho_f$, $\rho_h$, $w_k$, and $v_k$, their interaction with the objective must be further specified for the problem to be well posed. In this paper, we opt for a robust formulation in the min-max sense, where $\Theta^+_N$, the sequence of control parameters between $k$ and $k+N-1$, is chosen in such a way as to minimize $J_k$ in the worst-case realization of the uncertainties. The problem is also subjected to a set of $n_g$ inequality constraints
\begin{equation*}
    g\left(X^+_N,Y^+_N,U^+_N, W^+_N,V^+_N \right) \leq 0
\end{equation*}
which are required to hold across all possible realizations of the uncertainties, and with $W^+_N$ and $V^+_N$ defined similarly to $U^+_N$. A more complete mathematical formulation of this problem is presented in Section \ref{sec:out_feed} after the necessary tools have been introduced.
\begin{assumption}
    The function $g(\cdot,\cdot,\cdot,\cdot,\cdot):$ $\mathcal{X} \times \mathcal{Y} \times \mathcal{U} \times \mathcal{W} \times \mathcal{V} \rightarrow \mathbb{R}^{n_g}$ is continuous in all of its arguments, i.e.\  $g(\cdot,\cdot,\cdot,\cdot,\cdot) \in \mathcal{C}^{\,0}$.
\end{assumption}

Note that we make no assumptions on the convexity of $J_k(\cdot,\cdot,\cdot,\cdot,\cdot)$ or $g(\cdot,\cdot,\cdot,\cdot,\cdot)$, and leave the complexity of finding the global minima of these functions to the numerical solvers being employed, as discussed in Sections \ref{sec:out_feed} and \ref{sec:results}.

\section{Feasible State Space}
\label{sec:state}

In order to exploit our knowledge of the system dynamics of \eqref{eqn:dyn} in an output feedback formulation, we now proceed to describe the set of states that could be feasible for any given measurement sequence.

Let $\bar{w}_k := \{\rho_f,\rho_h,w_k,v_k\}$ be the collection of uncertainties that affect the system at time step $k$, and let $\bar{\mathcal{W}} := \mathcal{R}_f \times \mathcal{R}_h \times \mathcal{W} \times \mathcal{V}$ be the corresponding domain on which $\bar{w}_k$ is defined. Under Assumption \ref{assum:noise_compactness}, we know that $\bar{\mathcal{W}}$ must be compact. Given \eqref{eqn:dyn} and \eqref{eqn:measurements}, the set of states $\hat{\mathcal{X}}(Y,U)$ that are consistent with the dynamics and measurements $Y$ 
between time steps $k_1$ and $k_2$ (and with $U$ between $k_1$ and $k_2-1$) is described by
\begin{equation}
\begin{split}
    &\hat{\mathcal{X}}(Y,U) \coloneqq \left\{ X \left| \: \exists \bar{W} \in \bar{\mathcal{W}}: \: \begin{array}{l} \, \\ \, \end{array} \right. \right. \\
    & \left. \left.
    \begin{array}{ll}
        \forall i \in \{k_1,\hdots,k_2-1\}: & f(x_i,x_{i+1},u_i,\rho_f,w_i) = 0 \\
        \forall i \in \{k_1,\hdots,k_2\}: & y_i = h(x_i, u_i, \rho_h, v_i) 
    \end{array}
    \right. \right\}
\end{split}
\raisetag{46pt}
\end{equation}

For the sake of conciseness, we simply write $\hat{\mathcal{X}}$ to denote $\hat{\mathcal{X}}(Y,U)$ when the choice of $Y$ and $U$ are unambiguous. The set of possible states given measurements from time step $0$ to the current time step $k$ is therefore $\hat{\mathcal{X}}^-_k$, and $\hat{\mathcal{X}}^-_M$ is the set of possible states given measurements for the past $M < k$ time steps. By definition, $\hat{\mathcal{X}}^-_k$ exploits all of the available information on the true value of the state and is therefore the tightest possible description of the true value of $X^-_k$. However, this requires that all of $Y^-_k$ and $U^-_k$ be stored and is more expensive to compute than $\hat{\mathcal{X}}^-_M$. Consequently, it can be beneficial in some situations to condense the information obtained between time steps $0$ and $k-M-1$ into a set of inequalities $\beta \left( X^-_M \right) \leq 0$ constraining the possible values of $X^-_M$. To that end, we introduce the notation
\begin{equation}
    \left. 
    \hat{\mathcal{X}} \right| \beta \left( X \right) =  
     \hat{\mathcal{X}} \bigcap \left\{ X : \beta \left( X \right) \leq 0
     \right\}
\end{equation}
and propose approximating $\hat{\mathcal{X}}^-_k$ by $\hat{\mathcal{X}}^-_M|\beta \left( X^-_M \right)$ when previous measurements need to be discarded. 

One simple example of a set of inequalities $\beta \left( X^-_M \right)\leq 0$ can be obtained by determining a bounding box for each element in $x_{k-M}$ after computing its minimum and maximum values given $U$ and $Y$ between $0$ and $k-M-1$. However, it is important to note that this approach discards significant amounts of information compared to the full description of~$\hat{\mathcal{X}}^-_k$. To illustrate this, we consider the simple example with dynamics
\begin{subequations}
    \label{eqn:ex_sys_feasible}
\begin{align}
    x_{1,k+1} &= x_{1,k} + u_{1,k} + w_{1,k} \\
    x_{2,k+1} &= x_{1,k} + u_{1,k} + w_{1,k}
\end{align}
\end{subequations}
and measurement equations
\begin{subequations}
\begin{align}
    y_{1,k} &= x_{1,k} + v_{1,k} \\
    y_{2,k} &= x_{2,k} + v_{2,k}
\end{align}
\end{subequations}
where
\begin{subequations}
\begin{align}
    w_{1,k}^2 + w_{2,k}^2 &\leq 1 \\
    v_{1,k}^2 + v_{2,k}^2 &\leq 4.
\end{align}
\end{subequations}

\begin{figure}
    \centering
    \includegraphics[width = 0.7\columnwidth,trim={2cm 0 2cm 0},clip]{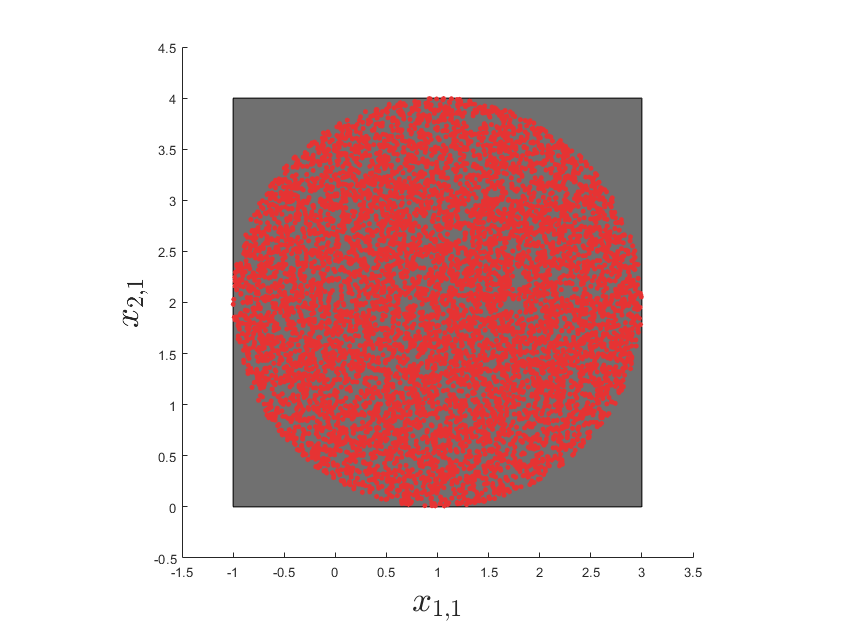}
    \caption{Comparison of the real feasible state space for $x_1$ (random samples drawn in red) to a box approximation (shown in grey) under the dynamics of \eqref{eqn:ex_sys_feasible} given the measurements $y_0 = [0,0]^\top$ and $y_1 = [1,2]^\top$, and the control inputs $u_0 = [1,1]^\top$.}
    \label{fig:box_v_true_uncertainty}
\end{figure}

We assume that the measurements $y_0 = \begin{bmatrix} 0 & 0\end{bmatrix}^\top$, and $y_1 = \begin{bmatrix} 1 & 2\end{bmatrix}^\top$ are available for the time steps $k = 0$ and $k = 1$, with the control input $u_0 = \begin{bmatrix} 1 & 1\end{bmatrix}^\top$ applied between the two time steps. Given these measurements, we know that $x_{1,1}$ must be between $-1$ and $3$, and that $x_{2,1}$ must be between $0$ and $4$, allowing us to build inequalities $\beta\left(x_1\right)$ restricting the elements of $x_1$ in these ranges. However, if we compute the true feasible space for $x_1$ through random samplings of solutions to $\hat{\mathcal{X}}^-_1$, as shown in Figure \ref{fig:box_v_true_uncertainty}, it quickly becomes evident that the real distribution is a circle, which is only a subset of the box approximation obtained from these lower and upper bounds. Thus, it is clear that the box constraints oversimplify the domain of $x_1$ and allow for otherwise impossible values of the state.

In general, $\hat{\mathcal{X}}^-_k$ can take arbitrary shapes that are potentially harder to describe than a simple circle. The use of longer measurement sequences (and therefore a larger $M$) in the definition of $\hat{\mathcal{X}}^-_M$ will therefore lead to a more realistic representation of the feasible state space at the expense of storage and computational costs. Some of the information from discarded measurements can be recovered through the constraints $\beta(\cdot)$, but doing so sacrifices accuracy compared to the original formulation.  


\section{Output Feedback}
\label{sec:out_feed}

Now that we can describe the set of possible states for our system given a set of measurements and inputs, we can proceed to formulate an optimization problem that solves for the optimal output feedback gains given a problem of the form described in Section \ref{sec:system}. To do so, we begin by solving for $J^*(\Theta^+_N)$, the worst-case value of the cost function given $\Theta^+_N$. This is described by the optimization problem
\begin{subequations}
\label{eqn:max_problem_output}
\begin{equation}
    J^*(\Theta^+_N) \coloneqq \max_{\substack{X, \, Y, \, U, \, \bar{W} }} J_k \left( X^+_N,\bar{X}^+_N,Y^+_N, \bar{Y}^+_N,U^+_N \right)
\end{equation}
such that
\begin{align}
    \label{eqn:w_constr_max}
    \bar{W} & \in \bar{\mathcal{W}} \\
    \label{eqn:state_feasible_max}
    X & \in \hat{\mathcal{X}} \left| \beta_k \left( X^-_M \right) \right. \\
    U^+_N & = \Phi_k(\Theta^+_N,Y) \\
    Y^-_M & = Y_0 \\
    \label{eqn:u0_constr_max}
    U^-_M & = U_0
\end{align}
\end{subequations}
where $Y_0$ represents the known measurements between $k-M$ and $k$, $U_0$ represents the control inputs applied between $k-M$ and $k-1$, and $\Phi_k(\Theta^+_N,Y)$ collects the output of $\phi_j(\theta_j,Y^{-}_{j,M})$ $\forall j \in [k,k+N-1]$, where $Y^{-}_{j,M}$ spans from $j-M$ to $j$. Note that in this context, \eqref{eqn:state_feasible_max} constrains the past states to agree with the observed measurements and forces future states and measurements to be compatible with each other. We also define the problem of finding the worst-case violation of any specific constraint $g_i$ as
\begin{equation}
\label{eqn:max_prob_g}
    g_i^*(\Theta^+_N) \coloneqq \max_{\substack{X, \, Y, \, U, \, \bar{W} }} \,
    g_i \left(X^+_N,Y^+_N,U^+_N,W^+_N,V^+_N \right)
\end{equation}
subject to the same constraints of \eqref{eqn:w_constr_max} through \eqref{eqn:u0_constr_max}.

Next, let $\mathcal{Z}\left(Y^-_M,U,\bar{W}\right)$ be the set of possible trajectories for $X$ and $Y^+_N$ given prior measurements $Y^-_M$ and specific trajectories of $U$ and $\bar{W}$ between $k-M$ and $k+N$, which can be defined as
\begin{equation}
\begin{split}
    \mathcal{Z}\left(Y^-_M,U,\bar{W}\right) \coloneqq & \left\{ X, Y^+_N \,  \left| \begin{array}{l} \, \\ \, \end{array} \right. \right. \\
    & \hspace{-50pt} \left.
    \begin{array}{l}
        \forall i \in \{k-M,\ldots,k+N-1\}: \\ \hspace{50pt} f(x_i,x_{i+1},u_i,\rho_f,w_i) = 0 \\
        \forall i \in \{k-M,\ldots,k\}: \\ \hspace{50pt} y_i = h(x_i, u_i, \rho_h, v_i) 
    \end{array} \hspace{-2pt}
    \right\}
\end{split}
\end{equation}
and let $\mathcal{Z}\left(Y^-_M,U,\bar{W}\right) \left| \beta_k \left( X \right) \right.$ be the similarly defined set that is compatible with the inequalities $\beta_k \left(X\right) \leq 0$.

The problem of designing an optimal robust output feedback policy in the sense of the formulation described in Section \ref{sec:system} can then be written as   

\begin{subequations}
\label{eq:semi_inf_output}
    \begin{equation}
    \min_{\Theta^+_N \, \in \, \varTheta} \; J^*(\Theta^+_N)
\end{equation}
s.t. $\forall \, \bar{W} \in \bar{\mathcal{W}}$, $\forall \left({X},Y^+_N\right) \in \mathcal{Z}(Y^-_M,U,\bar{W}) \left| \beta_k \left( X^-_M \right) \right.$,
\begin{equation}
\label{eqn:g_constr_min}
    g\left(X^+_N,Y^+_N,U^+_N,W^+_N,V^+_N \right) \leq 0
\end{equation}
where
\begin{align}
    U^+_N & = \Phi_k(\Theta^+_N,Y) \\
    Y^-_M & = Y_0 \\
    \label{eqn:u0_constr_min}
    U^-_M & = U_0.
\end{align}
\end{subequations}
\begin{assumption}
    \label{assum:theta_compact}
    The set $\varTheta$ of allowable parameter values for $\Theta^+_N$ is compact.
\end{assumption}
In \cite{wehbeh2024efficient}, it was shown that problems of the form of \eqref{eq:semi_inf_output} can be rewritten as semi-infinite programs of the form described by \cite{blankenship1976infinitely} despite the equality constraints for the dynamics and the implicit description of the feasible set $\mathcal{Z}$. Furthermore, it was shown that under the conditions of Assumptions \ref{assum:f_continuity} through \ref{assum:theta_compact}, and provided that the original formulation is feasible, these problems could be solved using local reduction. The method alternates between solving the problems of \eqref{eqn:max_problem_output} and \eqref{eqn:max_prob_g} to compute the worst-case scenarios for the cost and constraints given the current best guess for $\Theta^+_N$, and an approximation of the problem of $\eqref{eq:semi_inf_output}$ for which $\bar{\mathcal{W}}$ is replaced by the discrete scenario set $\bar{\mathbb{W}}_i$, which comprises the previous solutions to the problem of \eqref{eqn:max_problem_output} and \eqref{eqn:max_prob_g}. The algorithm then repeats this process until no uncertainties that lead to poorer performance or constraint violations can be found, returning optimal values of $\Theta^+_N$. Note, however, that in the case where the problem formulation is non-convex and local solvers are used to solve \eqref{eqn:max_problem_output}, \eqref{eqn:max_prob_g} and \eqref{eq:semi_inf_output}, the optimality of the solution returned cannot be guaranteed because the global maxima of the different problems cannot be determined in general. Instead, random restart or multi-start approaches must be used to increase the likelihood of the algorithm converging to the true global solution, which is required to guarantee robustness under the framerwork of~\cite{wehbeh2024efficient}.

\section{Simulation Results}
\label{sec:results}

\begin{figure*}[t]
\centering
\begin{subfigure}{.325\textwidth}
    \centering
    \includegraphics[width = \linewidth]{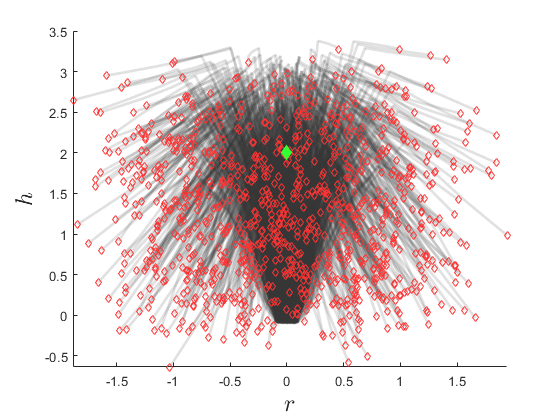}
    \caption{Open loop}
    \label{fig:open_loop}
\end{subfigure}
\begin{subfigure}{.325\textwidth}
    \centering
    \includegraphics[width = \linewidth]{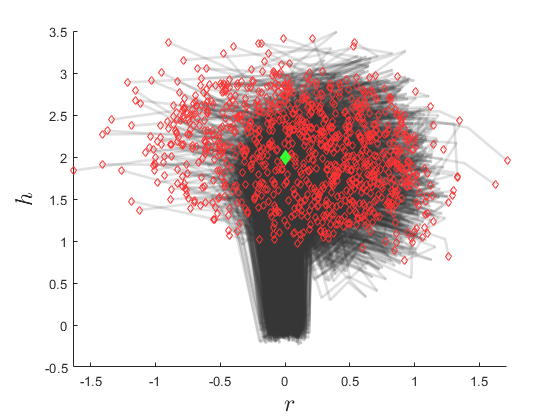}
    \caption{1-step controller}
    \label{fig:k1}
\end{subfigure}
\begin{subfigure}{.325\textwidth}
    \centering
    \includegraphics[width = \linewidth]{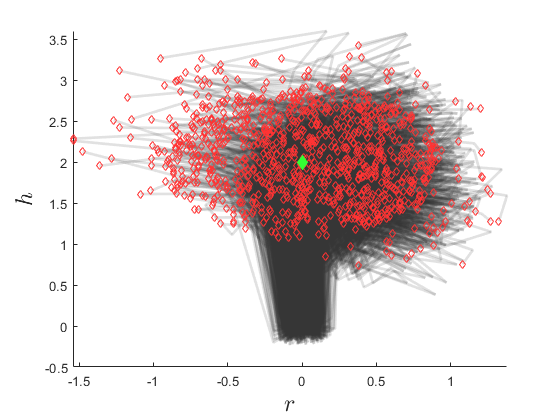}
    \caption{2-step controller}
    \label{fig:k2}   
\end{subfigure}
\vspace{3pt}
\caption{Sampled trajectories with end positions marked in red and the target marked in green.}
\label{fig:performance_runs}
\vspace{-8pt}
\end{figure*}

\begin{figure}[h]
    \centering
    \includegraphics[width=0.67\columnwidth]{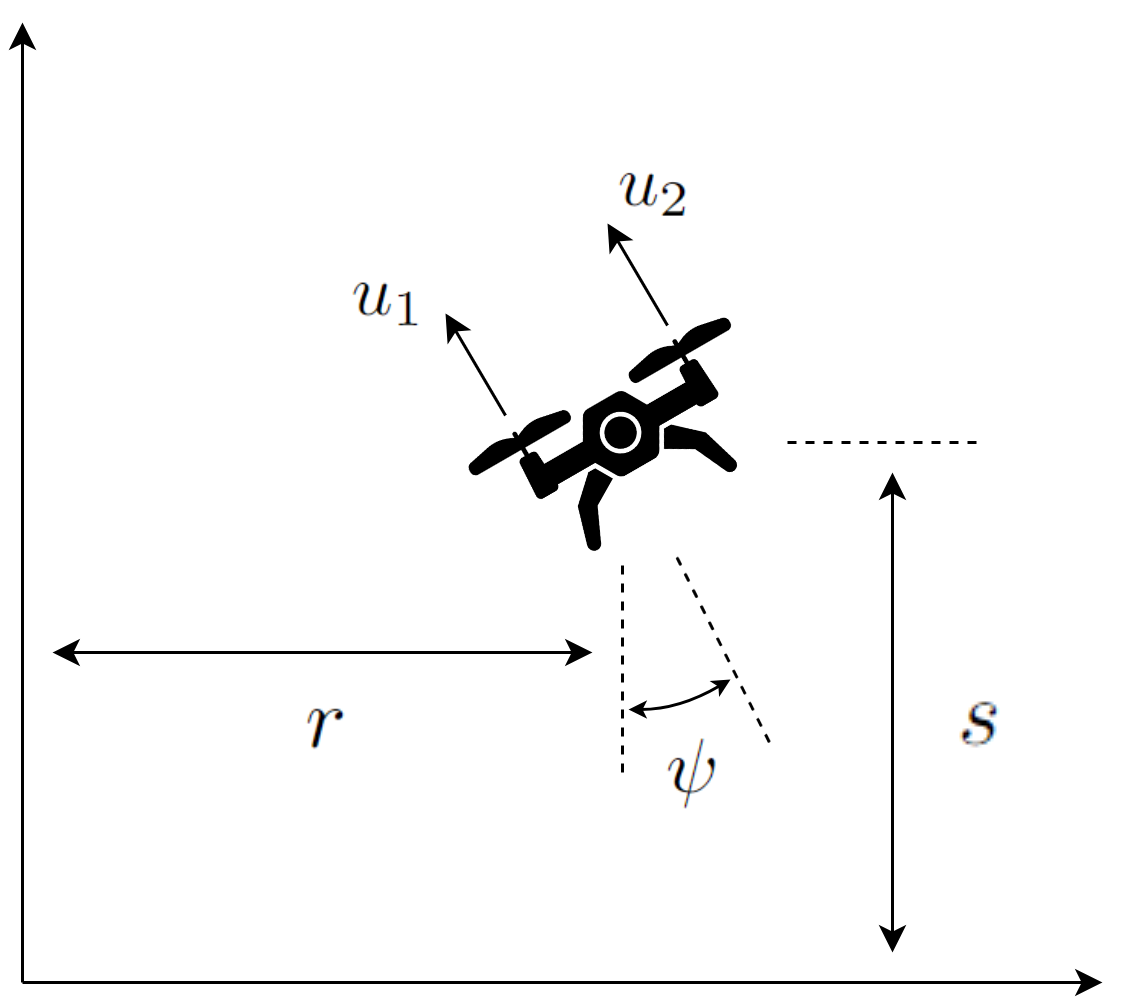}
    \caption{Illustration of the quadrotor states at time step $k$.}
    \label{fig:quad_example}
\end{figure}

In order to examine the performance of the control methodology of Section \ref{sec:out_feed} in practice, we now study the performance of three different control approaches on a trajectory tracking problem for a two-dimensional nonlinear quadrotor model with states $[r,\dot{r},s,\dot{s},\psi,\dot{\psi}]$, where $r$ is the quadrotor's horizontal position, $s$ is the quadrotor's height, and $\psi$ is the quadrotor's tilt angle, as illustrated in Figure \ref{fig:quad_example}.

The quadrotor's discrete-time dynamics, which are obtained through an Euler discretization with sampling time~$T_s$, can be expressed as
\begin{equation}
    \begin{bmatrix}
        {x}_{1,k+1} \\
        {x}_{2,k+1} \\
        {x}_{3,k+1} \\
        {x}_{4,k+1} \\
        {x}_{5,k+1} \\
        {x}_{6,k+1} 
    \end{bmatrix}
    =
    \begin{bmatrix}
        {x}_{1,k} \\
        {x}_{2,k} \\
        {x}_{3,k} \\
        {x}_{4,k} \\
        {x}_{5,k} \\
        {x}_{6,k} 
    \end{bmatrix} +
    T_s
    \begin{bmatrix}
    x_{2,k} \\[3pt]
    \frac{\sin(x_{5,k})\left(u_{1,k} + u_{2,k}\right)}{m}\\[3pt]
    x_{4,k} \\[3pt]
    \frac{\cos(x_{5,k})\left(u_{1,k} + u_{2,k}\right) }{m}- \gamma \\[3pt]
    x_{6,k} \\[3pt]
    \frac{\ell \left({u_{1,k}-u_{2,k}}\right)}{I}
    \end{bmatrix}
\end{equation}
where $x_{1,k}$ through $x_{6,k}$ are the discretized states at time $k$ corresponding to the continuous-time values $[r,\dot{r},s,\dot{s},\psi,\dot{\psi}]$, $0.9 \leq m \leq 1.1$ is the vehicle's mass, $0.001 \leq I \leq 0.0015$ is the moment of inertia, $\ell = 0.1$ is the moment arm for each motor, and $\gamma = 9.81$ is the gravity acting on the system. The measurement equation is then
\begin{equation}
    \begin{bmatrix}
        y_{1,k} \\
        y_{2,k} \\
        y_{3,k}
    \end{bmatrix}
    =
    \begin{bmatrix}
        x_{1,k} \\
        x_{3,k} \\
        x_{5,k}
    \end{bmatrix}
    +
    \begin{bmatrix}
        v_{1,k} \\
        v_{2,k} \\
        v_{3,k}
    \end{bmatrix}
\end{equation}
where $-0.1 \leq v_k \leq 0.1$ is the measurement noise at $k$. We examine the open loop control approach
\begin{equation}
    u_k = \bar{u}_k
\end{equation}
as well as the 1-step static  feedback law
\begin{equation}
    u_k = K_1 \, y_k + \bar{u}_k
\end{equation}
and the 2-step dynamic policy
\begin{equation}
    u_k = K_1 \, y_k + K_2 \, y_{k-1} + \bar{u}_k
\end{equation}
where $K_1$, $K_2$, and $\bar{u}_k$ are the decision variables that are to be optimized. 

The gain design problem is set up to minimize the worst-case of the objective
\begin{equation}
    J(x_{k+N}) = \left( x_{1,k+N} \right)^2 + \left( x_{3,k+N} - 2 \right)^2
\end{equation}
while guaranteeing that the constraint
\begin{equation}
    g(x_{k+N}) = x_{3,k+N} - 3.5 \leq 0 
\end{equation}
where $N=7$, $M=2$, all prior measurements are taken to be zero, and all prior control inputs are $4.905$. It is also known that $-0.05 \leq x_{2,k-M} \leq 0.05$.

For the sake of simplicity, we ignore thrust saturation, but note that it can be handled through the approach presented in \cite{wehbeh2024efficient}. Instead, we constrain the values of $K_1$ and $K_2$ to be between $-1.5$ and $1.5$ for the 2-step method, the values of $K_1$ to be between $-3$ and $3$ for the 1-step method, and the values of $\bar{u}$ to be between $-15$ and $15$ for the closed-loop methods, and between $-20$ and $20$ for the open-loop approach.

Simulations are carried out for the three control methods based on solutions obtained from an early termination of the local reduction method after the inclusion of 50 scenarios. The local reduction problems are implemented in Julia v1.10.2 using v0.19.0 of the JuMP package~\cite{Lubin2023} and v3.14.14 of the IPOPT optimizer~\cite{wachter2006implementation}, and run on a Microsoft Surface Studio laptop with an 11th Gen Intel\textsuperscript{\tiny\textregistered} Core\textsuperscript{\texttrademark} i7-11370H CPU at 3.30 GHz and 16GB of RAM. No solution requires longer than 50 seconds to compute, with the exact run times for each controller shown in Table \ref{tab:performance_metrics}.

The performance of the different methods is then evaluated based on 1000 different random mass, inertia and measurement noise values drawn uniformly from the intersection of $\bar{\mathcal{W}}$ with the set of uncertainties compatible with the previous measurements and dynamics, and used to obtain the trajectories plotted in Figure \ref{fig:performance_runs}. 

\begin{table}[h]
\centering
\caption{Key performance metrics for the different control approaches considered}
\label{tab:performance_metrics}
\begin{tabular}{@{}llll@{}}
\toprule
Control Method    & Avg. Cost & Max Cost & Runtime (s) \\ \midrule
Open-Loop & 1.901     & 8.015    & 16.41   \\
1-Step    & 0.619     & 3.01     & 37.34   \\
2-Step    & 0.545     & 2.746    & 44.28   \\ \bottomrule
\end{tabular}
\end{table}

The open-loop controller, shown in Figure \ref{fig:open_loop}, displays the poorest performance, failing to approach the target in a large number of cases. The addition of output feedback in the case of the 1-step controller shown in Figure \ref{fig:k1} leads to vastly improved performance, with a general clustering of the solutions in the vicinity of the target. Finally, the 2-step controller shown in Figure \ref{fig:k2} marginally improves on the 1-step approach by tightening the distribution around the target, and achieving better average and worst-case performance as seen in Table \ref{tab:performance_metrics}. None of the simulated trajectories for any of the control methods considered lead to any constraint violations, confirming the robustness of the gains produced.  


\section{Conclusion}
\label{sec:conclusion}

In this paper, we proposed a novel method to generate gains for arbitrary output feedback policies that can achieve robust performance on non-linear systems. The approach does not rely on any explicit state estimation but builds a set of all feasible states given knowledge of the dynamics, bounded uncertainties, and previous measurements and control inputs. A local reduction method is used to optimize the cost value associated with the worst-case prediction over this feasible state set, and ensure that none of the future feasible states violates any of the targeted constraints. This gain generation framework was then tested on a two-dimensional quadrotor control problem and achieved a good level of performance despite the uncertainty in the model description and the measurements given to the controller. Future work could aim to compare the performance of this method to approaches that rely on traditional state estimation, and quantify the potential benefits that can be achieved by considering the interactions between control inputs and future measurements. 

\addtolength{\textheight}{-0.5cm}   


\section*{ACKNOWLEDGMENTS}


\bibliographystyle{IEEEtran}
\bibliography{IEEEabrv,references.bib}

\end{document}